Thermoelectric properties of Co, Ir, and Os-Doped FeSi Alloys: Evidence for Strong Electron-Phonon Coupling

Brian C. Sales[1], Olivier Delaire[2], Michael A. McGuire[1] and Andrew F. May[1]

[1]Materials Sciences and Technology Division, Oak Ridge National Laboratory, Oak Ridge Tennessee, 37831

[2]Neutron Sciences Division, Oak Ridge National Laboratory, Oak Ridge Tennessee, 37831

**Abstract**

The effects of various transition metal dopants on the electrical and thermal transport properties of $Fe_{1-x}M_xSi$ alloys (M= Co, Ir, Os) are reported. The maximum thermoelectric figure of merit $ZT_{max}$ is improved from 0.007 at 60 K for pure FeSi to ZT = 0.08 at 100 K for 4% Ir doping. A comparison of the thermal conductivity data among Os, Ir and Co doped alloys indicates strong electron-phonon coupling in this compound. Because of this interaction, the common approximation of dividing the total thermal conductivity into independent electronic and lattice components ($\kappa_{Total} = \kappa_{electronic} + \kappa_{lattice}$) fails for these alloys. The effects of grain size on thermoelectric properties of $Fe_{0.96}Ir_{0.04}Si$ alloys are also reported. The thermal conductivity can be lowered by about 50% with little or no effect on the electrical resistivity or Seebeck coefficient. This results in $ZT_{max} = 0.125$ at 100 K, still about a factor of five too low for solid-state refrigeration applications.

**Introduction**

The compound FeSi is a narrow gap semiconductor with sharp peaks in the electronic density of states (DOS) at both the valence band and conduction band edges [1-7]. This generic type of electronic structure is believed to be advantageous for thermoelectric applications [8]. The present work explores the effects of various dopants on the electrical and thermal transport properties of FeSi. We find compelling evidence for strong electron-phonon coupling in this material.

FeSi is a member of a large class of interesting compounds (MnSi, CoSi, RuSi, OsSi, etc) that crystallize in the non-centrosymmetric cubic B20 structure (spacegroup $P2_13$) [9] (see Fig 1). Each iron is coordinated by 7 silicon atoms with 1 Si at 2.294 Å, 3 at 2.341 Å and 3 at 2.515 Å. The six next-nearest-neighbor Fe-Fe distances are each 2.753 Å, which are short enough to indicate significant direct Fe-Fe bonding. For example in iron metal the shortest Fe-Fe distance is 2.482 Å, and in $BaFe_2As_2$, a parent compound of the recently discovered layered iron superconductors, the distance is 2.80 Å [10]. In FeSi, hybridization between the iron 3d bands and Si 3p bands produces a small gap in the electronic DOS, and sharp peaks in the DOS near the gap edges result from large effective masses and accidental near degeneracies of multiple band extrema. Very similar features are found in the DOS of all of the transition metal silicides with the B20 structure (MnSi, CoSi, RuSi, OsSi, etc), although only for Fe, Ru, and Os silicide does the gap appear near the Fermi energy [3, 5, 11-13]. To account for the sharpness of the peaks near the gap edges in FeSi, correlations treated within the generalized gradient approximation (GGA) of density functional theory appear sufficient, although the calculated size of the hybridization gap is about a factor of 2 too large (see Fig 2). Estimates from optical and heat capacity data [14] yields an effective mass, m*, of about 30 times the free electron value for the carriers, a result qualitatively consistent with the flat bands near the gap edges, which result in the sharp peaks in the DOS (see Fig 2). Within this framework, if the effects of temperature and thermal disorder on both the electronic structure and the phonons are taken into account [5,15,19], the unusual temperature dependence of the optical conductivity [16, 17] and magnetic susceptibility can be explained without resorting to more exotic explanations. The focus of the present article, however, is to explore the effects of Ir, Co and Os doping on the electrical, thermal, and thermoelectric properties of FeSi. Of particular interest is the identification of novel approaches to improve ZT, the thermoelectric figure of merit where $Z = S^2/(\kappa\rho)$, and S is the Seebeck coefficient, $\kappa$ is the total thermal conductivity, and $\rho$ is electrical resistivity. Previous work [18] showed that Ir doping resulted in the largest value for ZT as compared to other electron dopants such as Co. Os doping was studied because it is isoelectronic with Fe (does not add carriers) and has approximately the same atomic mass as Ir. Thus the Os and Ir doped samples provide a way of separating the contributions of point-defect scattering and electron-phonon scattering to the thermal conductivity.

**Synthesis and Chemical Characterization**

Most of the samples were prepared by arc melting together high purity elements of Fe (99.99%), Si (99.999%), Ir (99.95 %), Co (99.99%) and Os (99.95%) on a water-cooled copper hearth in an argon atmosphere. If the samples were allowed to cool to room temperature after melting, the samples usually shattered upon further heating due to internal stress. To improve chemical homogeneity, after each melting the samples were flipped on the hearth plate while still hot ($\approx$ 600-800 C) and re-melted. Each sample was melted in this manner at least five times. A few of the arc-melted samples were annealed in vacuum at 1000 °C for 1 week to check if there was any significant change in properties due to improved chemical homogeneity. Our previous investigation of the effects of various dopants on the thermoelectric properties of FeSi found that Ir doping resulted in the best n-type thermoelectric performance of the dopants investigated [18]. Preliminary experiments in our laboratory established that an Ir concentration of about 0.04 ($Fe_{0.96}Ir_{0.04}Si$) is close to the optimum concentration (the concentration that resulted in the maximum value of ZT). Alloys with the same concentration of cobalt ($Fe_{0.96}Co_{0.04}Si$) and osmium ($Fe_{0.96}Os_{0.04}Si$) were prepared in order to compare the effects of another electron dopant (Co) or isoelectronic substitution (Os) on electrical and thermal transport. Powder x-ray diffraction (PXRD) measurements on all of the samples showed only the cubic B20 phase with refined lattice constants of 4.484 Å, 4.497 Å and 4.499 Å for 4% Co, Os and Ir doping, respectively. After annealing each of these samples in vacuum at 1000 °C, the lattice constants changed slightly to 4.485 Å, 4.500 Å and 4.499 Å. Energy dispersive x-ray spectroscopy (EDS) and scanning electron microscope (SEM) measurements in the backscatter mode indicated some spatial variation of the Ir and Os concentrations even after arc-melting multiple times and an 1000 °C anneal for 1 week. The concentration was uniform within 20 x 20 micron$^2$ regions but varied between 0.03 - 0.05 relative to iron over larger distances. The transport properties of a single crystal of FeSi, studied in [18] and [19], and an arc-melted FeSi sample were also measured in the present investigation for comparison purposes. The room temperature lattice constant for each sample determined from powder x-ray diffraction was 4.486 Å for both the single crystal and polycrystalline sample in good agreement with literature values [9]. Both FeSi samples had single-phase PXRD patterns, but EDS and SEM measurements from the arc-melted sample indicated about 1% of a ferromagnetic $Fe_3Si$ impurity phase.

Three polycrystalline samples of $Fe_{0.96}Ir_{0.04}Si$ were prepared with different crystalline grain sizes. One set of samples was cut with a diamond saw directly from the arc-melted sample or from an arc-melted sample after an additional anneal at 1000 °C in vacuum for 1 week. These polycrystalline samples had average grain sizes of 0.5-1 mm. The second set of samples was prepared by a coarse ball mill ($\approx$ 1-2h) of an arc-melted sample in an argon atmosphere. The powder produced via this process was passed through a 100 mesh sieve and had a wide distribution of particle sizes with grains typically in the 20-100 micron size with some grains much smaller (as estimated from SEM measurements). This powder was loaded into a graphite die in an argon atmosphere glove box and then quickly transferred to a spark-plasma-sintering (SPS) system and densified to near theoretical density. The typical conditions used for the SPS densification were a pressure of 25 MPa and about 1000 Amps of current through a 20 mm diameter sample. The third set of samples was prepared using a process similar to the second set. After the coarse ball milling, the powder was transferred in an argon glove box into a planetary mill and sealed. The powder was milled for an additional 40 h at 500 rpm. The resulting powder was extremely fine with most grains less than 0.1 micron. PXRD line widths suggested that a substantial fraction of this powder had grains of order 20 nm, but part of the broadening of the x-ray lines could be due to strain. This fine powder was loaded in a glove box into a graphite die and quickly transferred to the SPS system and densified. Samples for transport measurements, typically 10 x 2 x 2 $mm^3$, were cut from the fully dense samples using a low speed diamond saw.

**Experimental Methods**

Powder x-ray diffraction data were collected using a PANalytical X'Pert PRO MPD at room temperature using Cu Kα radiation. Scanning electron microscope and energy dispersive x-ray measurements were performed with a Hitachi TM-3000 tabletop microscope equipped with a Bruker Quantax 70 EDS system. Thermal conductivity, resistivity and thermopower data were collected from 300 to 2 K using the Thermal Transport Option (TTO) from Quantum Design and a 9 Tesla Physical Property Measurement system. Hall data were taken using a thin rectangular plate with typical dimensions of 8 x 4 x 0.4 mm$^3$ and a standard 4 lead geometry. Platinum wires (0.025 mm diameter) were attached to the samples using Epo-Tek H20E silver epoxy and Dupont 5790 silver paste. Heat capacity data were measured using the heat capacity option from Quantum Design.

**Results and Discussion**

The thermal conductivity, Seebeck coefficient and resistivity data are shown in Fig. 3 for two "pure" FeSi samples. Hall data at 2 K for each sample (not shown) indicate an extrinsic hole doping of about $10^{19}$ holes/cm$^3$ for the FeSi single crystal and $10^{18}$ holes/cm$^3$ for the polycrystalline FeSi sample. A small error in stoichiometry of order of $10^{-4}$ or $10^{-5}$, respectively, could account for this level of doping. The large peak in the Seebeck coefficient of FeSi near 35 K [20, 18] shown in Fig. 3b provided the initial motivation for investigating the potential of these materials for thermoelectric refrigeration. The temperature dependence and magnitude of the Seebeck coefficient are due to the unusual electronic DOS shown in Fig 2, as was first shown by Jarlborg (Ref. 5). Near the valence or conduction band edges, the density of states is highly asymmetric which produces a large magnitude for S with either light hole or electron doping. As the temperature is increased from 2 K up to about 50 K, S rapidly increases. At higher temperatures intrinsic electron-hole pairs are created and S decreases. At low temperatures the carrier concentration of the polycrystalline sample is about a factor of ten lower than the single crystal value. The corresponding larger value of S (1200 µV/K) is expected from standard semiconductor transport theory. What is surprising, however, is the factor of 2 increase in the thermal conductivity, κ, of the polycrystalline FeSi sample in the 20-70 K temperature range. In this temperature region the resistivity is high and it is expected that virtually all of the thermal conductivity is due to phonons. For example the estimated electronic contribution to κ from the Wiedemann-Franz relationship is less than 0.1 W/m-K at 50 K  There might be a small increase in κ due to reduced point defect scattering, but as noted above the variation in chemical stoichiometry is too small to produce a factor of 2 change in κ [21]. A likely origin of the reduction in κ is electron-phonon scattering. Similar reductions in κ occur when doping familiar semiconductors such as Si or SiGe alloys [22, 23]. To get a change in the thermal resistivity of Si at 50 K similar to that found for FeSi (see Fig 3a) requires a doping level of about $10^{21}$ carriers/cm$^3$ for Si [22], as compared to about $10^{19}$ carriers/cm$^3$ for FeSi.  Stronger electron-phonon scattering in FeSi is consistent with a much larger effective mass [14] for the carriers [23]. In the simplest models the strength of electron-phonon scattering is proportional to $m^{*2}$ [23].

For "pure" FeSi, the maximum value of ZT occurs near 60 K and is 0.007 and 0.013 for the polycrystalline and single crystal samples, respectively. To be useful for thermoelectric refrigeration, ZT should be at least about 0.6. To improve ZT the carrier concentration is usually manipulated through doping.  In general, however, it is difficult to predict the best way to dope a semiconductor [24]. Typically one uses the periodic table as a rough guide: one column to the right usually results in electron doping, while one column to the left results in hole doping. For example, in FeSi, replacing part of the Fe with Co, Rh or Ir should (and does, see ref. 18) result in a n-type semiconductor, while replacing part of the Si with Al, B, or Ga results in a p-type material. All dopants, however, are not equally effective, and it very difficult to predict *apriori* which dopant will work best. This is illustrated in Fig 4, where the effects of replacing 4% of the Fe by either Ir or Co are compared. The data for the Co doped sample are consistent with previous literature data [14, 25].  Both dopants result in an electron carrier concentration of about 2 x $10^{21}$ carriers/cm$^3$ as estimated from low temperature Hall measurements and the approximation that both Ir and Co add 1 electron carrier/atom. A similar carrier concentration for the Co doped sample was estimated from optical data [14]. These authors [14] were also able to estimate from optical and heat capacity data an effective mass of about 30 times the free electron value for the carriers. For all three properties the Ir dopant improves ZT relative to Co. With Ir doping the thermal conductivity and resistivity are lower and the Seebeck coefficient higher than when doping with Co (Fig 4). These data are all consistent with that reported previously [14, 18, 25]. A lower thermal conductivity with Ir doping is understandable since the mass difference between Ir and Fe is much larger than the difference between Fe and Co. This leads to much larger point defect scattering which is proportional to $(1-m_i/m_{av})^2$ where $m_i$ is the atomic mass of the dopant and $m_{av}$ is the average atomic mass of the alloy [21]. This factor is 80 times larger for Ir relative to Co doping. A lower electrical resistivity with Ir vs Co doping might qualitatively be understood since the 3d bands of Co are narrower than the 5d bands of Ir, and one might expect less magnetic carrier scattering with Ir relative to Co. This argument, however, does not explain the larger Seebeck coefficient for Ir (Fig 4b). First principles electronic structure calculations for an Ir doped alloy ($Ir_1Fe_{31}Si_{32}$, see Fig 2) or a Co doped alloy ($Co_1Fe_{31}Si_{32}$) indicate the Fermi energy is shifted into the peak at the bottom of the conduction band in good agreement with a rigid band shift. First principle calculations also indicate that replacing Fe with small amounts of either Ir or Os increases the size of the hybridization gap (Fig. 2) whereas for Co doping the gap is slightly smaller.

To further explore the effects of different dopants on the transport properties of FeSi alloys, we prepared an alloy with a 4% doping level of Os ($Fe_{0.96}Os_{0.04}Si$.). Os is in the same column of the periodic table as Fe, but has a mass within 1% of the Ir mass. The transport properties of this alloy are compared to the data from the FeSi single crystal and the 4% Ir alloy in Fig. 5. As expected, the resistivity of $Fe_{0.96}Os_{0.04}Si$ is similar to the FeSi single crystal since Os is isoelectronic with Fe and should not add carriers. Also the Seebeck coefficient for the Os doped sample (not shown) is very similar to FeSi data with a large and positive peak of 250 μV/K at 35 K. We compare the Os doped data to the FeSi single crystal data because these two samples have similar carrier concentrations at low temperature ($\approx 10^{19}$ carriers/cm$^3$). Osmium doping should reduce the lattice thermal conductivity due to point defect scattering, as is illustrated in Fig 5a. What is surprising is the thermal conductivity of the Ir doped sample. The reduction in the thermal conductivity of the lattice due to point defect scattering should be nearly the same for the Os and Ir doped samples and hence we expect $\kappa_{lattice\ Ir} \approx \kappa_{total\ Os}$. The additional carriers in the Ir sample should provide another channel for heat conduction ($\kappa_{electronic}$) yet below 150 K, the total thermal conductivity of the Ir doped sample is significantly *less* than that of the Os doped sample. This means that for the electrically conducting Ir doped sample, the common approximation of independent electronic and lattice components ($\kappa_{Total} \approx \kappa_{Lattice} + \kappa_{electronic}$) fails for this FeSi alloy. The same results were obtained on arc-melted samples annealed at 1000 C for 1 week. The strong coupling between the electrons and phonons in these alloys makes it impossible to cleanly separate the heat conduction into two distinct channels. The strong scattering of phonons by electrons in FeSi at low temperatures is likely related to the large effective mass ($\approx 30\ m_e$, Refs.14, 23) of the carriers. More quantitative measurements and calculations of the unusually strong interaction between the electrons and phonons in FeSi are given in [19].

The maximum ZT for the arc melted $Fe_{0.96}Ir_{0.04}Si$ alloy is ZT = 0.08 at T=90 K. The thermal conductivity at 90 K is 5 W/m-K, which is about a factor of ten higher than required for a good thermoelectric material in this temperature range. To reduce the lattice thermal conductivity, polycrystalline samples were prepared with different crystallite sizes as described in the synthesis section. The basic idea is to reduce the crystallite size so that the grain boundaries scatter the long wavelength acoustic phonons and hence reduce the thermal conductivity without increasing the electron scattering rate. This is possible if the electron mean free path is much shorter than the relevant phonon mean free paths [26]. Three different sets of samples were prepared and are labeled as "arc-melted", "ball milled", or 'planetary milled'. The arc-melted sample had large grains typically 0.5 –1mm in size. Both the ball milled and planetary milled powders had a large distribution of crystallite sizes with typical dimensions of 20-100 microns for the ball milled material and less than 0.1 microns for the planetary milled samples (see synthesis section). Transport data from the three $Fe_{0.96}Ir_{0.04}Si$ alloys are shown in Fig 6. The resistivity (Fig 6c) and Seebeck (Fig 6b) data from the three alloys are the same within our experimental error. The sample-to-sample variation in the measured value of the resistivity and Seebeck coefficient was below 5% and 2%, respectively. The thermal conductivity data (Fig 6a), however, exhibits a systematic decrease as the crystallite size is reduced. To verify this result, at least two different samples from different batches were measured for each type of sample (arc-melted, ball-milled, or planetary milled). The results shown in Fig 6 were reproduced. The values of ZT for the three types of $Fe_{0.96}Ir_{0.04}Si$ alloys are shown in Fig 7. Although there has not been a systematic optimization of the synthesis process, the planetary milled samples already show a 50% enhancement in ZT relative to the arc-melted alloys.

It is hypothesized that the increase in ZT is caused by crystallites that are smaller than the mean free paths of acoustic phonons that carry heat but larger than the electron mean free path. A rough estimate of the electron mean free path, $d_{electron} = 1.5\pi h/(e^2 k_F^2 \rho)$ where h is Planck's constant, e the electron charge, $\rho$ the resistivity and $k_F$ the Fermi wavevector [25]. Using the known carrier concentration ($\approx 2 \times 10^{21}$ electrons/cm$^3$) and resistivity at 90 K gives $d_{electron} \approx 3$ nm. Determining a mean free path for the phonons, $d_{phonons}$, is not as straightforward. The simplest expression for the lattice thermal conductivity is: $\kappa = 1/3\ C_v v_s d_{phonon}$, where $C_v$ is the heat capacity per unit volume and $v_s$ is an average sound velocity (about 3500 m/s for FeSi, Ref. 14 and Fig. 8). The heat capacity per unit volume, however, should only be the heat capacity of the acoustic phonons that carry significant amounts of heat. The calculated phonon dispersion curves for FeSi (which agree well with the measured dispersion curves [19] ) are shown in Fig 8. These data are also consistent with previous calculations [13]. The peak in ZT (Fig 7) is at about 90 K, which corresponds to an energy of about 8 meV. An advantage of investigating the effects of crystallite size on thermal conductivity at low temperatures is evident from Fig. 8. At 90 K (8 meV) the phonons excited are primarily one of the 3 acoustic branches. Near room temperature and above, a large fraction of the 21 optical branches are also excited. If the total measured heat capacity (not shown) at 90 K (7 J/mole-atoms-K) is used, however, the corresponding value for $d_{phonon}$ is only 4 nm, close to the estimated electron mean free path of 3nm. If, however, one uses the measured or calculated phonon density of states, $g(\omega)$, [19] and assume that only acoustic phonons with energies less than a cutoff value contribute to heat transport [26], the value $C_v$ is much less. From Fig 8, a cutoff value of about 20 meV avoids the mixing of the acoustic and optic modes and results in a heat capacity of 1.65 J/mole-atoms-K at 90 K, which results in $d_{phonon} \approx$ 17 nm. If a cutoff of 8 meV is used $d_{phonon} \approx$ 150 nm. The uncertainty in the estimates of the relevant phonon mean free paths is a direct consequence of our lack of understanding as to how much heat is carried by acoustic phonons with different wavelengths. A better microscopic understanding of heat transport in solids is required [see ref. 26 and references therein].

**Summary and Conclusions**

The effects of various transition metal dopants on the electrical and thermal transport properties of $Fe_{1-x}M_xSi$ alloys (M= Co, Ir, Os) are presented. The maximum thermoelectric figure of merit ZT is improved from 0.007 at 60 K for pure FeSi to ZT = 0.08 at 100 K for 4% Ir doping. A comparison of the thermal conductivity data among Os, Ir and Co doped alloys indicates strong electron-phonon coupling in this compound since adding electrons, and hence another channel for heat conduction, results in a lower total thermal conductivity. The common approximation of dividing the total thermal conductivity into independent electronic and lattice components ($\kappa_{Total} = \kappa_{electronic} + \kappa_{lattice}$) fails for these alloys. As carriers are added to FeSi, strong electron-phonon scattering significantly reduces the heat conducted by phonons. This means that $\kappa_{lattice}$ is a strong function of the carrier concentration and is not an independent quantity. Compared to doped Si [22], the reduction is much stronger presumably due to the large effective mass of the carriers ($\approx 30\ m_e$) in FeSi [14, 23]. The effects of small crystallite size on the thermoelectric properties of $Fe_{0.96}Ir_{0.04}Si$ alloys are also reported. It is found that the thermal conductivity can be lowered by about 50% with little or no effect on the electrical resistivity or Seebeck coefficient. This results in $ZT_{max} = 0.125$ at 100 K, still about a factor of five too low for solid-state refrigeration applications. The processing parameters used in reducing the crystallite grain sizes have not been optimized and further improvements in ZT are likely. The estimated electron mean free path in these alloys is 3nm. The estimated phonon mean free path, however, varies from 4nm to 150 nm depending on the assumptions used in the estimate. A better microscopic understanding of heat transport in real compounds is needed, however, to provide a more rational guide to the use of grain size in the design of better thermoelectric materials.

**Acknowledgements**

It is a pleasure to acknowledge useful discussions with David Mandrus, David Parker, David Singh, and Paul Kent. The technical assistance of Hu Longmire, Larry Walker and Ed Kenik is gratefully acknowledged. Research sponsored by the Material Sciences and Engineering Division, Office of Basic Energy Sciences, U. S. Department of Energy. Early portions of this research were supported by the ORNL LDRD program. O. D. was sponsored partially by the Scientific Users Facilities Division, Office of Basic Energy Sciences U.S. DOE and a D.O.E. Frontier Research Center, DE-SC00001299.

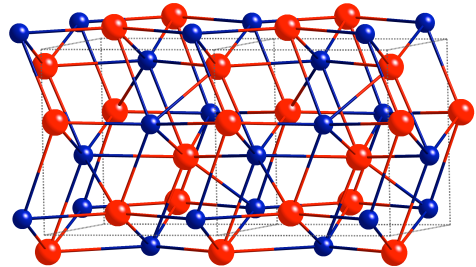

Figure 1 (color online) Crystal structure of FeSi with Fe (red, larger sphere) and Si (blue, smaller sphere). Each Fe is coordinated by 7 Si atoms and each Si is coordinated by 7 Fe. The shortest Fe-Fe distance is only 2.75 Å, which suggests direct Fe-Fe bonding. The dotted lines outline the unit cell.

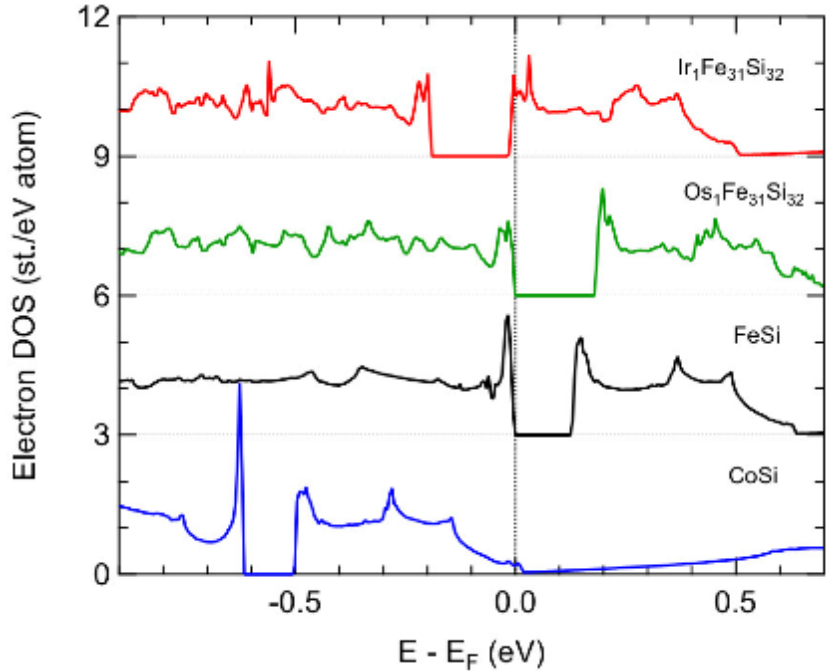

Figure 2 (color online) Electronic density of states for FeSi, CoSi and two FeSi alloys ($Ir_1Fe_{31}Si_{32}$ and $Os_1Fe_{31}Si_{32}$) calculated from first principles. Correlations were included using the PBE-96 generalized gradient exchange correlation functional. Note that hybridization between the transition metal d bands and the Si p bands produces a small gap in the density of states with sharp peaks in the density of states at the gap edges. For FeSi and doped alloys this gap appears near the Fermi energy, $E_F = 0$. For clarity the DOS curves for each composition are offset along the vertical axis.

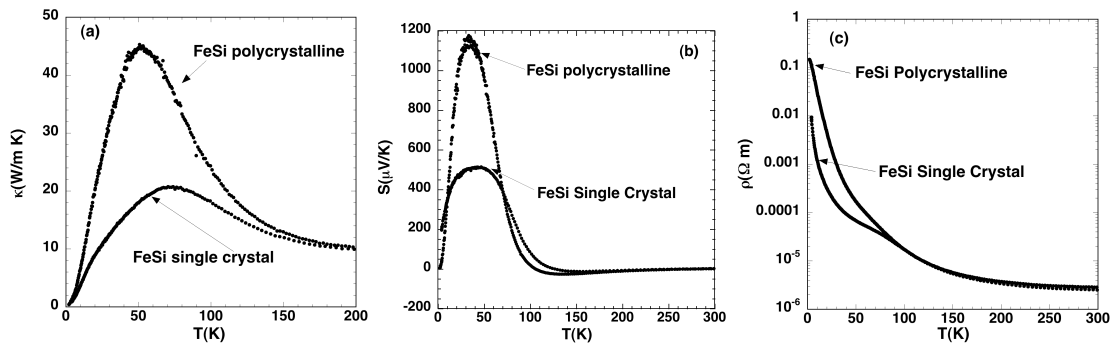

Fig 3 Transport data [(a) thermal conductivity, (b) Seebeck and (c) resistivity] from a FeSi single crystal and a polycrystalline FeSi sample with about a factor of 10 lower extrinsic carrier concentration. The "knee" in the resistivity curve at about 70 K is more prominent in some FeSi samples than in others and likely depends on the specific origin of "doping" in these samples. There is no phase transition associated with this feature.

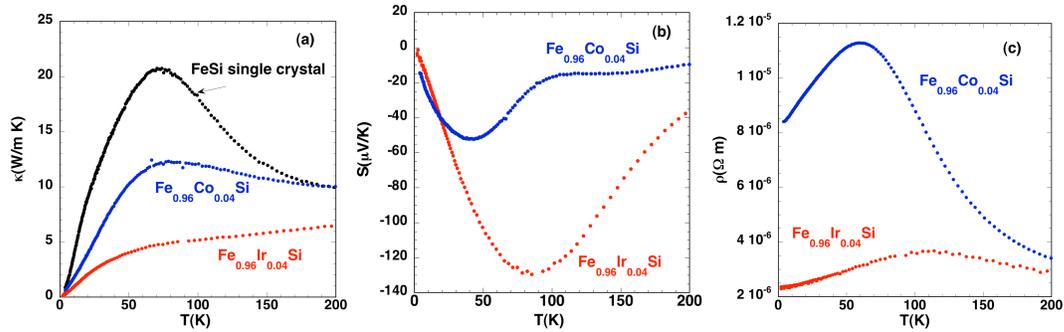

Fig 4 (Color online) A comparison of the effects of 4% Co or Ir doping on (a) thermal conductivity (b) Seebeck coefficient and (c) electrical resistivity. Doping with Ir is better than Co for thermoelectric performance at 100 K for all three properties.

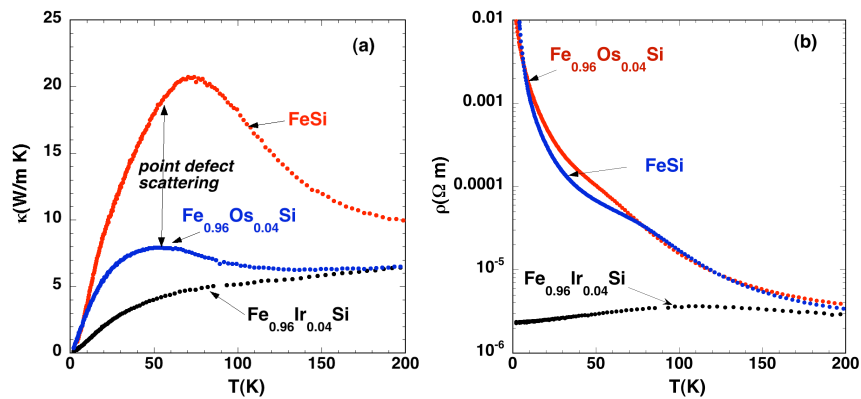

Fig 5 (Color online) Comparison of the effects of 4% Ir or 4% Os doping on the (a) thermal conductivity and (b) resistivity. Os and Ir have similar atomic masses (within 1%) which should result in similar reductions in the lattice thermal conductivity due to point defect scattering. Os is isoelectronic to Fe and does not add carriers with doping unlike Ir which adds approximately 1 electron/Ir atom. Note, however, that the total thermal conductivity of the Ir doped sample below 150 K is *less* than that of the Os doped sample due to strong electron-phonon scattering (see text).

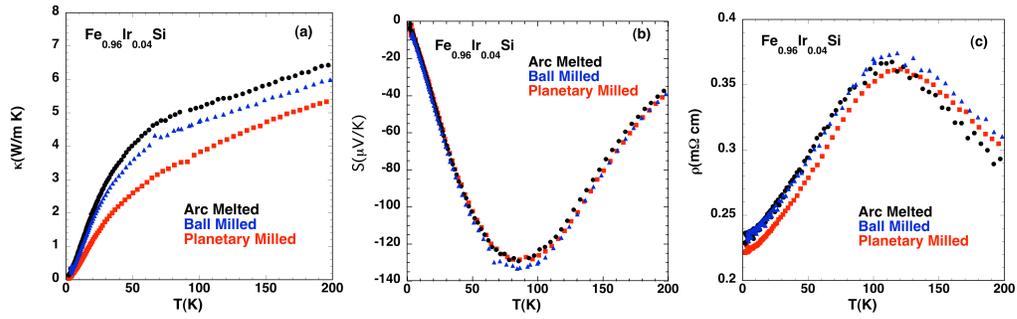

Fig 6. (Color online) Effects of crystallite size on the (a) thermal conductivity, (b) Seebeck coefficient and (c) electrical resistivity of a 4% Ir doped FeSi alloy. (circles, arc melted largest grains; triangles, ball milled smaller grains; squares, planetary milled smallest grains)

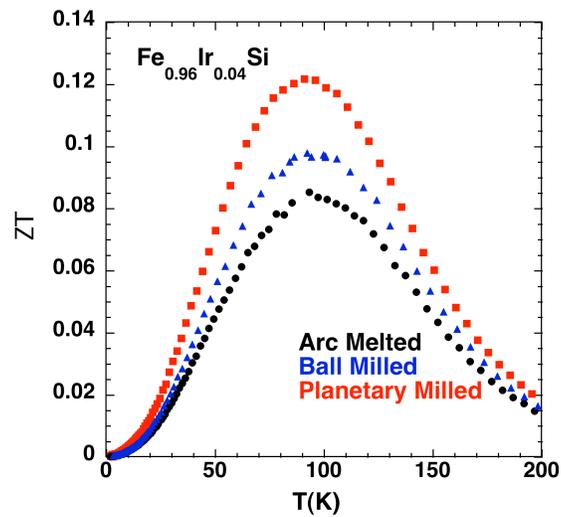

Fig 7 (Color online) ZT versus T for a 4% Ir doped FeSi alloy. (circles, arc melted largest grains; triangles, ball milled smaller grains; squares, planetary milled smallest grains) (See text for details)

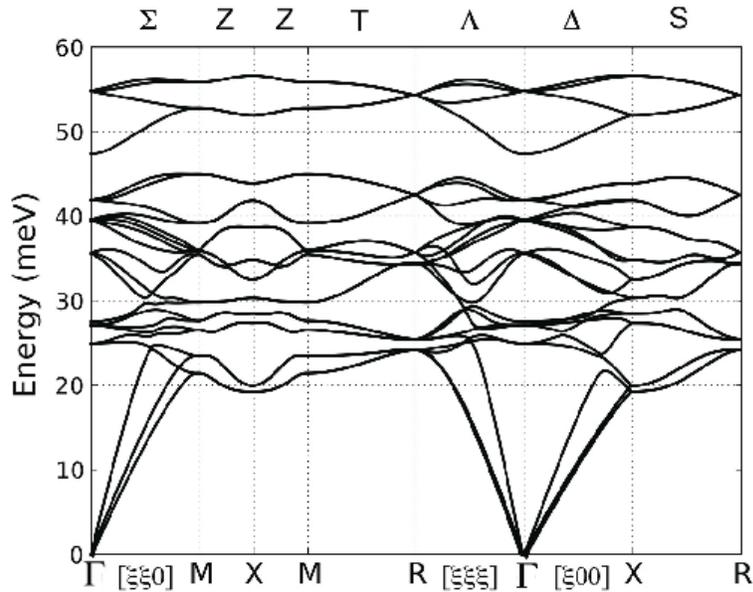

Fig 8 Phonon dispersion curves for FeSi calculated from first principles (24 branches: 21 optical and 3 acoustic). Calculated dispersion curves are in good agreement with the dispersion curves measured using inelastic neutron scattering [19].